\documentstyle[12pt]{article}

\begin{document}
{\sf \begin{center} \noindent {\Large\bf Slow dynamos and decay of monopole plasma magnetic fields in the Early Universe}\\[2mm]

by \\[0.1cm]

{\sl  L.C. Garcia de Andrade}\\Departamento de F\'{\i}sica
Te\'orica -- IF -- Universidade do Estado do Rio de Janeiro-UERJ\\[-3mm]
Rua S\~ao Francisco Xavier, 524\\[-3mm]
Cep 20550-003, Maracan\~a, Rio de Janeiro, RJ, Brasil\\[-3mm]
Electronic mail address: garcia@dft.if.uerj.br\\[-3mm]
\vspace{1cm} {\bf Abstract}
\end{center}
\paragraph*{}
Previously Liao and Shuryak [\textbf{Phys. Rev C (2008)}] have
investigated electrical flux tubes in monopole plasmas, where
magnetic fields are non-solenoidal in quark-QCD plasmas. In this
paper slow dynamos in diffusive plasma [{\textbf{Phys. Plasmas
\textbf{15} (2008)}}] filaments (thin tubes) are obtained in the
case of monopole plasmas. In the absence of diffusion the magnetic
field decays in the Early Universe. The torsion is highly chaotic in
dissipative large scale dynamos in the presence of magnetic
monopoles. The magnetic field is given by the Heaviside step
function in order to represent the non-uniform stretching of the
dynamo filament. These results are obtained outside the junction
condition. Stringent limits to the monopole flux were found by Lewis
et al [\textbf{Phys Rev D (2000)}] by using the dispute between the
dynamo action and monopole flux. Since magnetic monopoles flow
dispute the dynamo action, it seems reasonable that their presence
leads to a slow dynamo action in the best hypothesis or a decay of
the magnetic field. Hindmarsh et al have computed the magnetic
energy decay in the early universe as ${E}_{M}\approx{t^{-0.5}}$,
while in our slow dynamo case linearization of the growth rate leads
to a variation od magnetic energy of ${\delta}{E}_{M}\approx{t}$,
due to the presence of magnetic monopoles. Da Rios equations of
vortex filaments are used to place constraints on the geometry of
monopole plasma filaments.{\bf PACS
numbers:\hfill\parbox[t]{14.5cm}{Plasma dynamos,47.65.Md, Magnetic
monopoles 14.80.Hv.}}}
\newpage
 \section{Introduction}
 As early as $1983$ Arons and Blanford \cite{1} have shown that dynamo growth of galactic field can still occur in the scales of $10^{7}-10^{8}yr$ in the presence of a monopole halo.
 Resonant character of magnetic field damping by moving monopoles ,allows the field to survive indefinetely when monopole plasma frequency is large enough.
 More recently Liao and Shuryak \cite{2} have investigated electric flux tubes in monopole plasmas, in order to better understand the strong quark-gluon
 plasma (QCD). Earlier, Brandenberg and Zhang \cite{3} have linked galactic dynamo mechanism to QCD scale, by using string (thin flux tubes) dynamics to explain galactic
magnetic fields by dynamo action \cite{4}. Earlier Lewis et al
\cite{5} have investigate this dispute between monopole dissipation
and dynamo action in protogalactic ambient. They found more
stringent limits to the monopole flux than the ones found by Parker.
In this report it is shown that by considering a Riemannian model
\cite{6} of curved stretch-twist and fold \cite{7} magnetic dynamo
flux tube \cite{8}, the monopole current is created at the junction
condition of the flux tube. These Riemannian tubes used here are
similar to the Moebius strip, considered by Shukurov et al \cite{9}
to model experiments such as the Perm torus dynamo \cite{10}.
Earlier, Hindmarsh et al \cite{11} were able to show that the
magnetic field decay energy is given by $E_{M}\approx{t^{-0.5}}$
\cite{12} in the early universe. The model we shall be discussing in
this session is a slow dynamo model in the presence of dissipation
given by the monopole flux. Here , we show that the magnetic energy
decay is given by $E_{M}\approx{t}$ , which is higher, due to the
presence of the magnetic monopoles. This paper is organized as
follow: In section 2 the dynamo filament model of non-uniform
stretching as thin channels, for the monopole plasmas is presented
and in section 3 the solution of self-induction equation is given
and the physical properties of the dynamo action in monopole plasma
universe is obtained. In this section, the Da Rios equation are
shown to place constraints in the torsion and curvature of plasma
monopole filaments. In Section 4, finally addresses conclusions.
 \newpage
\section{Monopole plasma dynamo flow}
\vspace{1cm} Let us start this section, by defining the main
equations of the self-induction and the monopole ansatz magnetic
field. By defining the magnetic field as
\begin{equation}
\textbf{B}= B_{0}(t)H_{0}(s-s_{0})\textbf{t} \label{1}
\end{equation}
where $H_{0}$ is the Heaviside step function which physically means
that along the filament coordinate s, the junction condition
establish the point where the magnetic field is turned on. The
stationary flow speed is defined by
\begin{equation}
\textbf{v}:={v}_{0}{\delta}(s-s_{0})\textbf{t} \label{2}
\end{equation}
where $v_{0}$ is constant and ${\delta}(s-s_{0})$ is the delta Dirac
distribution. When $s_{0}=nT$ where $n \in{\textbf{Z}}$ is an
integer, this flow is a chaotic flow \cite{13}, with period T. Here
${\textbf{t}}$ is the tangent vector to the plasma filament, based
on the Frenet frame $(\textbf{t},\textbf{n},\textbf{b})$, which
describes the time evolution. The self-induction equation
\begin{equation}
{\partial}_{t}\textbf{B}={\nabla}{\times}(\textbf{v}{\times}\textbf{B})+\eta{\nabla}^{2}\textbf{B}\label{3}
\end{equation}
since both flow speed and magnetic field are along the magnetic flow
as in the London equation ${\textbf{J}}= {\lambda}\textbf{B}$
\cite{14}. Magnetic monopoles are characterized by the violation of
the solenoidal equation, which is replaced by the monopole equation
\begin{equation}
{\nabla}.\textbf{B}:={\rho}_{m} \label{4}
\end{equation}
where ${\rho}_{m}$ is the magnetic monopole density. By Substitution
of equation (\ref{4}) into the equation (\ref{3}) one obtains the
scalar equation
\begin{equation}
{\partial}_{t}{\rho}_{m}=\eta{\nabla}^{2}{\rho}_{m}\label{5}
\end{equation}
where the first term on the RHS of equation vanishes which shows by
simple inspection that the monopole density decays in time,
physically showing why the monopoles are so difficult to find in the
present universe. Nevertheless as show by Arnon and Blanford , the
monopole density can be very high in galactic scales where dynamo
growth can recover this loss. The model we shall be discussing in
this session is a slow dynamo model in the presence of dissipation
given by the monopole flux. Now let us discussed the model based on
plasma monopole filaments where the dynamo action, is given by
stretching the filament, which in turn produces a non-uniform
stretching which is fundamental for a strong magnetic field. The
presence of magnetic monopole flux seems to explain the reason why a
fast dynamo action does not take place and in its place a slow
dynamo shows up. At the junction condition the plasma flow is
compressible, in the sense that the divergence of the flow is
\begin{equation}
{\nabla}.\textbf{v}=v_{0}\frac{{\delta}(s-s_{0})}{[s-s_{0}]}
\label{6}
\end{equation}
which diverges at the junction $s=s_{0}$, note however that, the
integration of this equation yields
\begin{equation}
\int{({\nabla}.\textbf{v})dV}=\frac{v_{0}{\pi}r^{2}}{[s-s_{0}]}
\label{7}
\end{equation}
which shows that the integral does not necessarilly diverge at the
junction, since the radius of the filament $r\rightarrow{0}$, and so
is the magnetic field away from the junction condition, namely the
magnetic monopoles are absent away from the junction condition of
plasma filament. Note that off the junction this integral vanishes
and the flow is incompressible away from the junction. Therefore
there is a flow of monopoles at the junction. Monopole density as
high as $10^{18}\frac{GeV}{c^{2}}$ in the galactic halo. In our case
monopole density diverges and need to be normalized  by a quantum
theory. The density ${\rho}_{m}$ is given by
\begin{equation}
{B}_{0}{\delta}(s-s_{0}):={\rho}_{m} \label{8}
\end{equation}
By assuming that $B_{0}=e^{{\gamma}t}$, where ${\gamma}$ is the
growth rate of the magnetic field, substitution of the magnetic
field ansatz into the self-induction equation yields
\begin{equation}
[{\gamma}+{\delta}(s-s_{0})]\textbf{B}-{\kappa}(s,t)B_{0}[{\tau}(s,t)+v_{0}{\delta}(s-s_{0})]H_{0}\textbf{n}+B_{0}{\kappa}_{s}H_{0}\textbf{b}={\eta}[B_{0}{\kappa}_{s}H_{0}\textbf{n}+2B_{0}{\kappa}{\delta}\textbf{n}+{\delta}'
\frac{1}{{H}_{0}}\textbf{B}] \label{9}
\end{equation}
where one has used the following Frenet generalized equations
\cite{14}
\begin{equation}
\frac{{d}\textbf{t}}{dt}=-{\kappa}(s,t){\tau}(s,t)\textbf{n}+{\kappa}_{s}\textbf{b}
\label{10}
\end{equation}
\begin{equation}
\frac{{d}\textbf{t}}{ds}={\kappa}(s,t)\textbf{n}
 \label{11}
\end{equation}
\begin{equation}
\frac{{d}\textbf{n}}{ds}=-{\kappa}(s,t)\textbf{n}+\tau\textbf{b}
\label{12}
\end{equation}
and
\begin{equation}
\frac{{d}\textbf{b}}{ds}=-{\tau}(s,t)\textbf{n} \label{13}
\end{equation}
Where here ${\kappa}$ and $\tau$ represents respectively Frenet
curvature and torsion, of the plasma filaments. Expression (\ref{9})
can be further simplified to
\begin{equation}
{\gamma}_{eff}\textbf{B}-{\kappa}(s,t)B_{0}[{\tau}(s,t)+v_{0}{\delta}(s-s_{0})]H_{0}\textbf{n}+B_{0}{\kappa}_{s}H_{0}\textbf{b}={\eta}[B_{0}({\kappa}_{s}
H_{0}+2{\kappa}{\delta})\textbf{n}+B_{0}{\kappa}_{s}{{H}_{0}}\textbf{b}]
\label{14}
\end{equation}
where the effective growth rate now is given by
\begin{equation}
{\gamma}_{eff}={\gamma}+{\delta}-{\kappa}^{2} \label{15}
\end{equation}
which shows that curvature of filaments appears non-linearly in the
dynamo equation for the monopole dynamo filamentation. One notes
therefore that the curvature of plasma filamentation contributes to
weaken dynamo effects while the presence of Dirac delta function
acts as to enhance the dynamo action.
\section{Slow dynamos and constraints in plasma filamentation curvature and torsion}
In this section we shall solve the self-induction equation and apply
the results to found when dynamos or decay or magnetic fields takes
place in the early universe under monopole structures. Separation of
the equation (\ref{14}) along the Frenet frame yields the following
three equations
\begin{equation}
{\gamma}=-{\delta}+{\eta}[{\kappa}^{2}+\frac{{\delta}}{[s-s_{0}]H_{0}}]
\label{16}
\end{equation}
\begin{equation}
{\kappa}{\tau}=\frac{v_{0}}{2}{\delta}(s-s_{0}) \label{17}
\end{equation}
and
\begin{equation}
{\kappa}_{s}={\eta}{\tau}{\kappa} \label{18}
\end{equation}
From the first equation (\ref{16}) in this group it is clear that in
the diffusion free limit ${\eta}\rightarrow{0}$, the dynamo growth
${\gamma}$ vanishes in the absence of monopoles, or away from the
junction of plasma filaments, where the Dirac function vanishes upon
integration. Thus this actually implies that in the absence of
monopoles the dynamo is slow, while in the presence of monopoles the
dynamo effect is absent in the diffusion or dissipative case, since
then
\begin{equation}
{\gamma}=-{\delta}(s-s_{0}) \label{19}
\end{equation}
which represents a strong and very fast decay of monopole magnetic
fields in the junction of plasma filaments. The remaining equations
can be used along with the Da Rios vortex filaments equations
\begin{equation}
{\kappa}_{t}=-2{\kappa}_{s}{\tau}-{\tau}_{s}\kappa\label{20}
\end{equation}
and
\begin{equation}
{\tau}_{t}=[-\frac{{\kappa}_{ss}}{\kappa}+\frac{{\kappa}^{2}}{2}-{\tau}^{2}]_{s}
 \label{21}
\end{equation}
By assuming that the plasma filaments are stationary or
${\kappa}_{t}={\tau}_{t}=0$, yields
\begin{equation}
{\tau}(s)=-\frac{1}{{\eta}s}
 \label{22}
\end{equation}
which shows that the dissipation brought by monopole current yields
a damping on the torsion of the filamentation process. The Frenet
curvature is finally given by
\begin{equation}
{\kappa}^{2}=\frac{1}{4{\eta}{s}^{2}}
 \label{23}
\end{equation}
The magnetic field is expressed as
\begin{equation}
\textbf{B}=e^{{\gamma}t}H_{0}(s-s_{0})\textbf{t}
 \label{24}
\end{equation}
while the repeated process that leads to the dynamo chaotic
structure can be obtained by simply substitution of $s_{0}=nT$.
Computation of the magnetic energy integral
\begin{equation}
E_{M}=\frac{1}{8{\pi}}\int{\textbf{B}^{2}dV}=\frac{1}{8{\pi}}\int{e^{{\gamma}t}H_{0}(s-s_{0})\textbf{t}dV}
 \label{25}
\end{equation}
which yields ${\delta}E_{M}\approx{{\eta}{\kappa}^{2}t}$ obtained by
linearizing the exponential stretching. By considering equations
(\ref{17}), (\ref{22}) and (\ref{23}) yields
\begin{equation}
\frac{1}{4{\eta}^{2}s^{2}}=\frac{v_{0}}{2}{\delta}(s-s_{0})
 \label{26}
\end{equation}
which by integration and applying the Dirac constraint
\begin{equation}
\int{{\delta}(s-s_{0})ds}=1
 \label{27}
\end{equation}
at the junction condition $s=s_{0}$, one obtains
\begin{equation}
\int{\frac{1}{4{\eta}^{2}{s}^{2}}ds}=\frac{v_{0}}{2}\int{{\delta}(s-s_{0})ds}
 \label{28}
\end{equation}
or
\begin{equation}
v_{0}=-\frac{1}{4{\eta}^{2}{s_{0}}}
 \label{29}
\end{equation}
which completely determines the constant velocity of the plasma
dynamo flow in terms of the dissipation. Since the diffusion
coefficient ${\eta}$ is inversely proportional to the magnetic
Reynolds number $Re_{m}$ one obtains finally that
\begin{equation}
v_{0}=-\frac{{Re_{m}}^{2}}{4{s_{0}}}
 \label{30}
\end{equation}
which shows that the plasma filamentary flow is damped in small
scale dynamos where $R_{m}$ are small and is enhance for large scale
astrophysical and cosmic dynamos. This is physically interesting
since in astrophysical dynamo flows the speed of plasma monopoles
should be faster than in non-monopole plasmas in laboratory.
\newpage

\section{Conclusions}
Plasma filamentation which is used in this brief report in the
context of the monopole plasma in cosmology, may also be used in
tokamak and plasma instabilities \cite{15}. Earlier Uby et al
\cite{16} have used also, Frenet frame plasma filamentation dynamics
to investigate the effects on the superconductivity type II which
makes explicity use of London equations. The results obtained here
reinforce the existence of dynamo action versus monopole dissipation
in the early universe, giving rise to an energy which is
proportional to ${\delta}E_{M}\approx{t}$ where the proportionally
coefficient is given by the product of dissipative coefficient
$\eta$ and the curvature of plasma filaments.
\section{Acknowledgements}:
\newline I am very much in debt to Dmitry Sokoloff for helpful discussions on the subject of this
paper. Financial supports from Universidade do Estado do Rio de
Janeiro (UERJ) and CNPq (Brazilian Ministry of Science and
Technology) are highly appreciated.

\end{document}